\documentclass[pra,aps,twocolumn,showpacs]{revtex4}
\usepackage{graphicx,graphics,epsfig}
\usepackage{dcolumn}
\usepackage{bm}
\usepackage{amsmath}
\usepackage{verbatim}
\usepackage{color}
\usepackage[colorlinks=false]{hyperref} 
\usepackage{subfigure}
\usepackage{times,natbib}
\usepackage{amsmath,amsfonts,amssymb,graphics,graphicx,epsfig,color,times,natbib}
\usepackage{tikz}
\usepackage{pgfplots}
\usepackage{verbatim}
\usepackage{fancyhdr} 
\newcommand{\tr}{\mbox{tr}}
\newcommand{\Exp}{\mbox{exp}}

\newcommand{\Un}{\raisebox{-0.52\height}{\includegraphics[height=0.8cm]{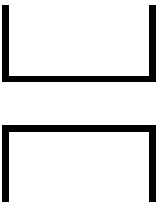}}}
\newcommand{\Unij}{\raisebox{-0.52\height}{\includegraphics[height=0.8cm]{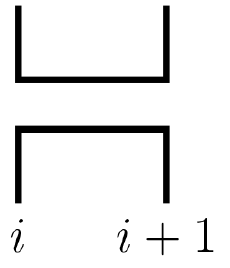}}}
\newcommand{\llll}{\raisebox{-0.52\height}{\includegraphics[height=0.8cm]{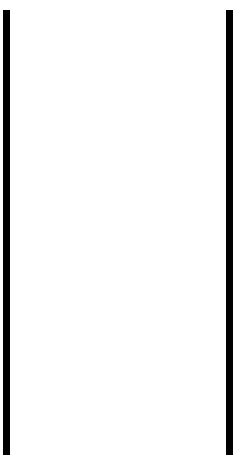}}}
\newcommand{\cross}{\raisebox{-0.52\height}{\includegraphics[height=0.8cm]{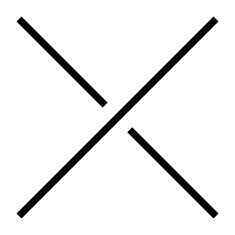}}}
\newcommand{\myloop}{\raisebox{-0.4\height}{\includegraphics[height=0.5cm]{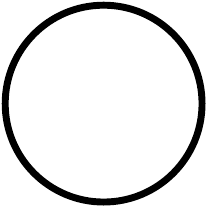}}}
\newcommand{\UNUN}{\raisebox{-0.4\height}{\includegraphics[height=1.2cm]{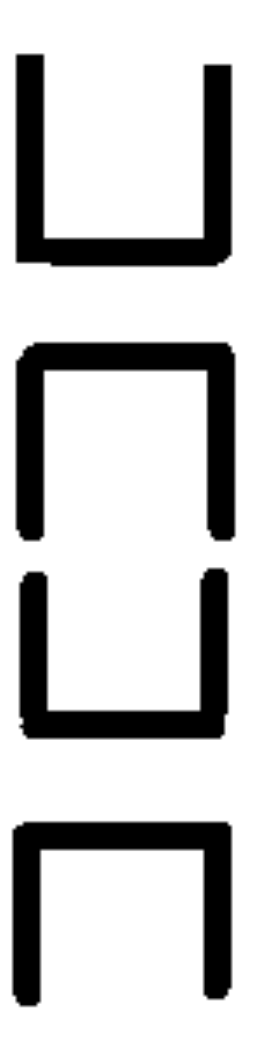}}}
\newcommand{\ununun}{\raisebox{-0.4\height}{\includegraphics[height=1.2cm]{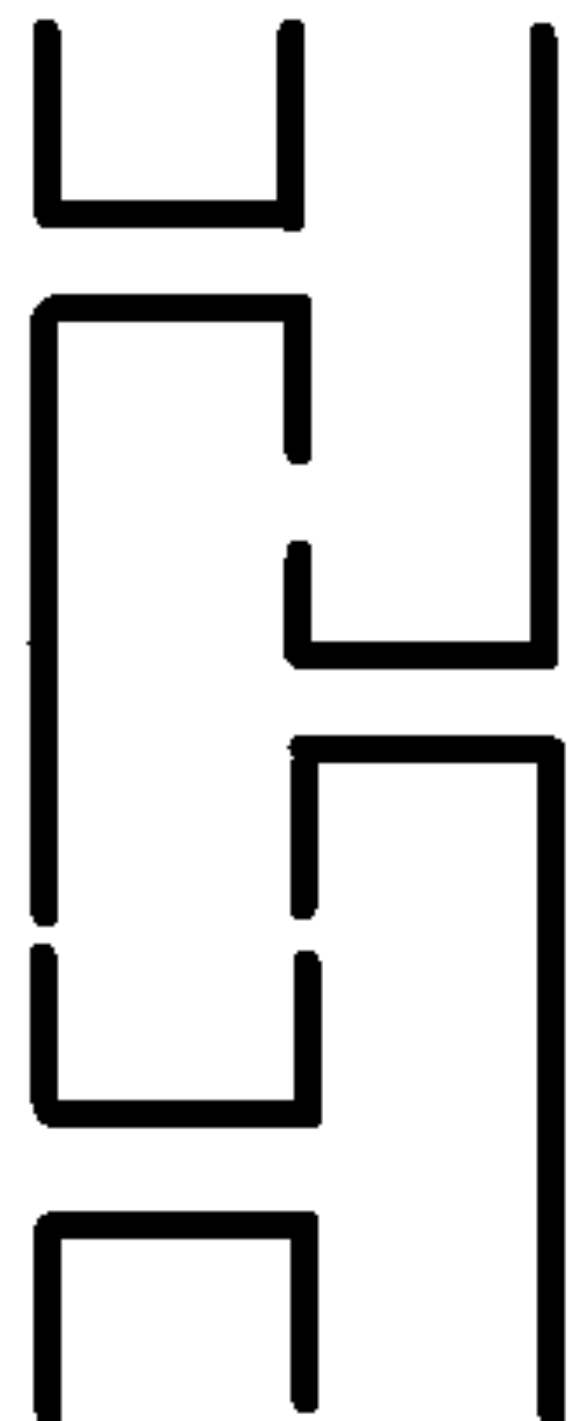}}}
\newcommand{\unL}{\raisebox{-0.4\height}{\includegraphics[height=1.2cm]{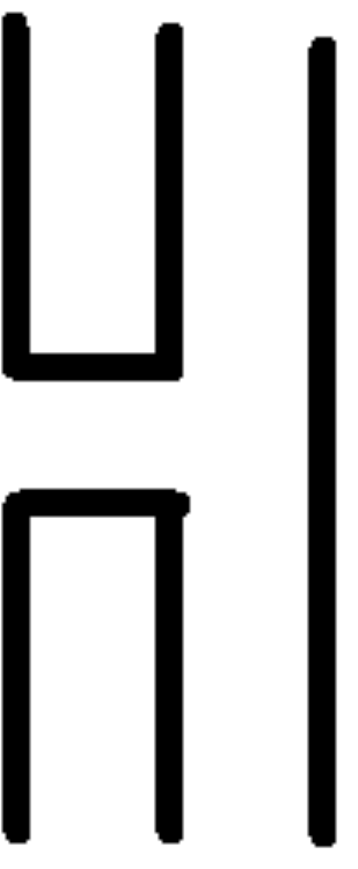}}}
\newcommand{\UON}{\raisebox{-0.4\height}{\includegraphics[height=1.2cm]{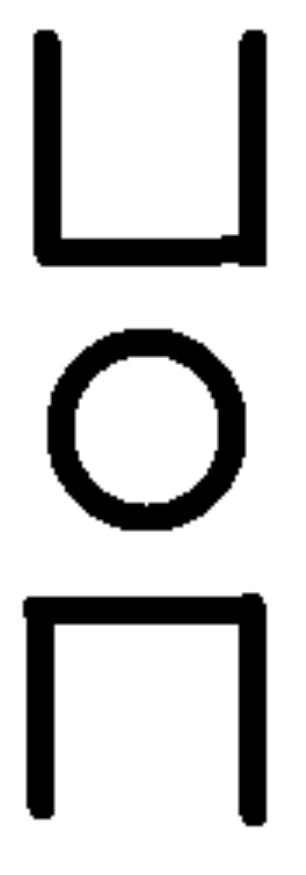}}}
\newcommand{\smyloop}{\raisebox{-0.4\height}{\includegraphics[height=0.4cm]{loop}}}
\newcommand{\Uu}{\raisebox{-0.4\height}{\includegraphics[height=0.6cm]{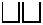}}}
\newcommand{\twojoin}{\raisebox{-0.4\height}{\includegraphics[height=0.6cm]{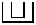}}}

\begin{document}
\title{$\ell_1$-norm and  entanglement in screening out braiding from\\Yang-Baxter equation associated with $\mathbb{Z}_3$ parafermion}
\author{Li-Wei Yu}
\email{nkyulw@yahoo.com}
\affiliation{Theoretical Physics Division, Chern Institute of Mathematics, Nankai University, Tianjin 300071, China}

\author{Mo-Lin Ge}
\email{geml@nankai.edu.cn}
\affiliation{Theoretical Physics Division, Chern Institute of Mathematics, Nankai University, Tianjin 300071, China}

\date{\today}


\begin{abstract}

\textbf{The relationships between quantum entangled states and braid matrices have been well studied in recent years. However, most of the results are based on qubits.  In this paper, We investigate the applications of  2-qutrit entanglement in the braiding associated with $\mathbb{Z}_3$ parafermion. The 2-qutrit entangled state $|\Psi(\theta)\rangle$, generated by acting the localized unitary solution $\breve{R}(\theta)$ of YBE on 2-qutrit natural basis,  achieves its maximal $\ell_1$-norm and maximal von Neumann entropy simultaneously at $\theta=\pi/3$.  Meanwhile, at $\theta=\pi/3$, the solutions of YBE reduces braid matrices, which implies the role of $\ell_1$-norm and entropy plays in determining real physical quantities. On the other hand, we give a new realization of 4-anyon topological basis by qutrit entangled states, then the $9\times9$ localized braid representation in 4-qutrit tensor product space $(\mathbb{C}^3)^{\otimes 4}$ are reduced to Jones representation of braiding in the 4-anyon topological basis. Hence, we conclude that the entangled states are powerful tools in analysing the characteristics of braiding and $\breve{R}$-matrix.}
\end{abstract}
\maketitle

\section{Introduction}\label{Intro}
Since Kauffman and Lomonaco pointed out the relations between quantum entanglement and topological entanglement  \cite{kauffman2002quantum,kauffman2004braiding}, the $4\times4$ braid matrix and Yang-Baxter equation(YBE) have been widely applied to entanglement and quantum information from various aspects  \cite{ge2012yang}. For example, the solution of YBE was used for describing 2-qubit entanglement \cite{zhang2005universal,chen2007braiding} and 3-qubit entanglement \cite{yu2014factorized}, the entanglement was connected with Berry phase in Yang-Baxter system \cite{chen2007braiding,chen2008berry,wang2009entanglement,yu2014factorized}, quantum optical realization of YBE \cite{hu2008optical}, entanglement in describing quantum criticality of YBE chain model \cite{hu2008exact} and so on. On the contrary, sometimes the entanglement can also be employed in obtaining braiding out of Yang-Baxter equation(YBE).  One of which is to detect the role of extremum of $\ell_1$-norm and von Neumann entropy of qubit entangled state in determining the $4\times4$ braid matrix from the  parametrized  $\breve{R}(\theta)$-matrix which satisfies Yang-Baxter equation \cite{niu2011role}. In previous works, the relations between quantum entanglement and localized unitary representation of braiding as well as $\breve{R}$-matrix have been well discussed. However, most of the results are limited in the qubit entanglement and $4\times4$ braid matrix. Recently, based on the new localized $D^2\times D^2$ representation of braid group associated with $\mathbb{Z}_{D}$ parafermion \cite{rowell2012localization,hastings2013metaplectic,wang2014multipartite}, the relations between qudit entanglement and braiding have been discussed \cite{wang2014multipartite}. Due to the potential applications of the $\mathbb{Z}_D$ parafermionic type of topological quantum models in quantum computing and quantum information \cite{hutter2016quantum}, it is valuable to investigate further the relations between qudit entanglement and $D^2\times D^2$ braiding as well as YBE.

Now we briefly introduce some results about $4\times4$ YBE in quantum information. Different from the original 6-vertex and 8-vertex $4\times4$ localized unitary solutions of YBE originating from chain models, in N-qubit tensor product space $(\mathbb{C}_2)^{\otimes N}$,  the new type of $4\times4$ localized solution $\breve{R}$-matrix associated with quantum information reads ($I$ is $2\times2$ identity matrix) \cite{ge2012yang}
\begin{eqnarray}
&&\breve{R}_{i}(\theta_1) \breve{R}_{i+1}(\theta_2) \breve{R}_{i}(\theta_3)=\breve{R}_{i+1}(\theta_3) \breve{R}_{i}(\theta_2) \breve{R}_{i}(\theta_1),\label{YBEcondition}\\
&& \breve{R}_i(\theta)=I\otimes I\cdots\otimes \underset{i,i+1}{ \breve{R}(\theta)}\otimes\cdots I\otimes I,\\
&&   \breve{R}(\theta) = \left[
   \begin{array}{cccc}
   \cos\theta & 0 & \ 0 & \ \sin\theta  \\
   0 & \cos\theta & \ \sin\theta & \ 0 \\
   0 & -\sin\theta & \ \cos\theta & \ 0 \\
   -\sin\theta  & 0 & \ 0 & \ \cos\theta
   \end{array} \right]\label{4solution}.
\end{eqnarray}
Eq. (\ref{YBEcondition}) is the so-called Yang-Baxter equation. 
Under the solution shown in Eq. (\ref{4solution}), the constraint for parameters in Eq. (\ref{YBEcondition}) obeys the Lorentzian type additivity $\tan\theta_2=\frac{\tan\theta_1+\tan\theta_3}{1+\tan\theta_1\tan\theta_3}$, whereas the original 6-vertex solutions obey the Galilean type additivity $u_2=u_1+u_3$. Acting $\breve{R}$-matrix on 2-qubit natural basis $\{|00\rangle,|01\rangle, |10\rangle, |11\rangle\}$, where $|00\rangle=(1,0,0,0)^T,|01\rangle=(0,1,0,0)^T,|10\rangle=(0,0,1,0)^T,|11\rangle=(0,0,0,1)^T$, one obtains four 2-qubit pure states. Without loss of generality, we choose one state
\begin{equation}
|\psi(\theta)\rangle=\cos\theta|00\rangle-\sin\theta|11\rangle.
\end{equation}
Clearly, the parameter $\theta$ in $\breve{R}$  describes the entangled degree of $|\psi(\theta)\rangle$. When $\theta=\pi/4$, $|\psi(\theta)\rangle$ turns to be the maximal entangled state, meanwhile, the $\breve{R}$-matrix becomes braid matrix associated with $SU(2)_2$ Chern-Simons theory,
\begin{equation}
  B = \frac{1}{\sqrt{2}}\left[
   \begin{array}{cccc}
   1 & 0 & \ 0 & \ 1  \\
   0 & 1 & \ 1 & \ 0 \\
   0 & -1 & \ 1 & \ 0 \\
   -1  & 0 & \ 0 & \ 1
   \end{array} \right],
\end{equation}
i.e. the braid matrix describes the maximal entangled state. In comparison with braid $B$, the advantage of $\breve{R}(\theta)$ is that the $\theta$ describes any entangled degree of 2-qubit pure states. On the other hand,  from the viewpoint of $\ell_1$-norm, the extremum of $\ell_1$-norm of $|\psi(\theta)\rangle$ exactly corresponds to the maximal entanglement as well as the braid matrix\cite{niu2011role}. Hence, $\ell_1$-norm endows the real physical meaning for $\breve{R}(\theta)$-matrix: the braiding operation. 

In this paper,  we mainly focus on the role of $\ell_1$-norm and 2-qutrit entanglement playing in braiding representation associated with $\mathbb{Z}_3$ parafermions, which are related to metaplectic anyon models in $SO(3)_2=SU(2)_4$ Chern-Simons(CS) theory \cite{hastings2013metaplectic,cui2015universal}. Firstly, based on the $\mathbb{Z}_3$ parafermion representation of Yang-Baxter equation, we investigate the application of $\ell_1$-norm and von Neumann entropy of the 2-qutrit state  generated by $\breve{R}$-matrix, and find that the extreme values of $\ell_1$-norm and von Neumann entropy correspond to the braiding. The result is the generalization of qubit case proposed in Ref. \cite{niu2011role}. Secondly, inspired by the qubit representation of 4-anyon basis for $SU(2)_2$ CS associated with $4\times4$ braid matrix \cite{niu2011role}, the two 4-anyon topological fusion basis in $SO(3)_2=SU(2)_4$ CS are represented by 4-qutrit entangled states. Then the well known $9\times9$ localized unitary representation of braiding is reduced to the Jones representation of braiding for 4-strand topological basis. 

The applications of $\ell_1$-norm in quantum physics have already been proposed in recent years, such as in quantum process tomography \cite{kosut2008quantum}, Yang-Baxter equation \cite{niu2011role} and quantifying coherence \cite{baumgratz2014quantifying}, {\em et al.}  In the previous work \cite{niu2011role}, the authors have shown the motivation of adopting $\ell_1$-norm in real physical model associated with $SU(2)_2$ CS. Here we give a brief introduction about it. Usually in quantum mechanics, a wave function $|\Phi\rangle$ can be expanded as $|\Phi\rangle=\sum_i\alpha_i|\phi_i\rangle$, where $|\phi_i\rangle$ is orthonormal basis. The normalization of $|\Phi\rangle$ reads
\begin{equation}
\left\langle\Phi|\Phi\right\rangle=\sum_i|\alpha_i|^2=1.
\end{equation}  
We call $\sum_i|\alpha_i|^2=||\alpha||_{\ell_2}$ as $\ell_2$-norm, which indicates the square integrability or the probability distribution of the wave function.  Meanwhile, $\ell_1$-norm is defined as 
\begin{equation}\label{l1norm}
||\alpha||_{\ell_1}=\sum_i|\alpha_i|.
\end{equation}
 The minimization of $\ell_1$-norm plays important roles in information theory such as Compressed Sensing theory \cite{donoho2006compressed,candes2006near}, {\em et al.} Hence, it is worthy of investigating whether the extremization of $\ell_1$-norm can be used to determine some important physical quantities in quantum information or in quantum mechanics. In Ref. \cite{niu2011role}, the extremum of $\ell_1$-norm has been well connected to $4\times4$ braid matrix, we now extend the results to $9\times9$ braid matrix. 

The paper is organized as follows. In Sec. \ref{Sec2}, we review the solutions of Yang-Baxter equation expressed by $\mathbb{Z}_3$ parafermion. In Sec. \ref{Sec3}, the $\ell_1$-norm and von Neumann entropy of the entangled state $|\Psi(\theta)\rangle$ generated by $\breve{R}$-matrix is discussed. In Sec. \ref{Sec4}, the $9\times9$ braid matrix are reduced to $2\times2$ Jones representation of braiding under the 4-strand topological basis represented by 4-qutrit entangled states. In the last section, we make the conclusion and discussion.

\section{Review of the parafermionic representation of braid operators and YBE}\label{Sec2}
The metaplectic anyons  have shown their universal topological quantum computation abilities under the braiding and measurement \cite{cui2015universal,levaillant2016topological}, hence it is valuable to study the characteristics of the type of anyons. In this section, we review the $\mathbb{Z}_3$ parafermionic representation \cite{rowell2012localization,hastings2013metaplectic,hutter2016quantum} of braid operators with quantum dimension $d=\sqrt{3}$, which is associated with metaplectic anyons in $SO(3)_2=SU(2)_4$ topological quantum field theory.  Based on the $SU(3)$ matrix representation of parafermions in tensor product space $(\mathbb{C}_3)^{\otimes N}$, the localized $9\times9$ unitary representation for braid relation (also known as Yang-Baxter equation without parameter) can be obtained. Moreover, the rational Yang-Baxterization \cite{yu2016z3} of the corresponding braid operators are shown below. 

We first introduce the $\mathbb{Z}_3$ parafermion.  As a natural generalization of Clifford algebra for Majorana fermion, the algebraic relation of $\mathbb{Z}_3$ parafermion $C_i$ reads
\begin{equation}\label{z3para}
\begin{aligned}
&C_iC_j=\omega^{\textrm{sgn}|j-i|}C_jC_i, \quad \omega=\Exp(\text{i}\frac{2\pi}{3}),\\
&(C_i)^2=C_i^{\dagger},\, (C_i^{\dagger})^2=C_i,\\
&(C_i)^3=(C_i^{\dagger})^3=1.
\end{aligned}
\end{equation}
Then the Temperley-Lieb(T-L) elements associated with braid relation are composed by nearest neighbor parafermions
\begin{equation}
T_i=\frac{1}{\sqrt{3}}\left(1+\omega^2C_i^{\dagger}C_{i+1}+\omega^2C_iC_{i+1}^{\dagger}\right),
\end{equation}
which satisfies Temperley-Lieb algebra \cite{temperley1971relations} for quantum dimension $d=\sqrt{3}$,
\begin{equation}
\begin{split}
&T_i^2=dT_i,\, d=\sqrt{3},\\
&T_iT_{i\pm1}T_i=T_i,\\
&T_iT_j=T_jT_i, \,|i-j|\leq1.
\end{split}
\end{equation}
The braid operator is expressed as 
\begin{equation}\label{BTL}
B_i=\omega(e^{-i\frac{\pi}{6}}T_i-1),
\end{equation}
and satisfies braid relation \cite{kauffman2001knots}
\begin{equation}
B_iB_{i+1}B_i=B_{i+1}B_iB_{i+1}.
\end{equation}
By defining the concrete matrix representation of $\mathbb{Z}_3$ parafermion in N-qutrit space $(\mathbb{C}_3)^{\otimes N}$ through ($I$ is $3\times3$ identity matrix) \cite{yu2016z3} 
\begin{equation}\label{z3jw}
\begin{aligned}
&C_{\textrm{2k-1}}^{\dagger}=Z^{\otimes (k-1)}\otimes X_{\textrm{k}}\otimes I^{\otimes (N-k)},\\
&C_{\textrm{2k-1}}=(Z^{\dagger})^{\otimes (k-1)}\otimes X^{\dagger}_{\textrm{k}}\otimes I^{\otimes (N-k)},\\
&C_{\textrm{2k}}^{\dagger}=Z^{\otimes (k-1)}\otimes(XZ)_{\textrm{k}}\otimes I^{\otimes (N-k)},\\
&C_{\textrm{2k}}=(Z^{\dagger})^{\otimes (k-1)}\otimes(XZ)^{\dagger}_{\textrm{k}}\otimes I^{\otimes (N-k)},
\end{aligned}
\end{equation}
where 
\begin{equation*}
Z=\left[\begin{array}{ccc} 1 & 0 & 0 \\ 0 & \omega & 0 \\ 0 & 0 & \omega^2 \end{array}\right], \, X=\left[\begin{array}{ccc} 0 & 1 & 0 \\ 0 & 0 & 1 \\ 1 & 0 & 0 \end{array}\right],
\end{equation*}
one can obtain the braid matrix from three nearest parafermionic sites in 2-qutrit space $(\mathbb{C}_3)^{\otimes 2}$:
\begin{eqnarray}
&&B_{12}=B_{1}=\left[\begin{array}{ccc} e^{-i\frac{\pi}{3}} & 0 & 0 \\ 0 &e^{i\frac{\pi}{3}} & 0 \\ 0 & 0 &e^{-i\frac{\pi}{3}} \end{array}\right]\otimes\left[\begin{array}{ccc} 1 & 0 & 0 \\ 0 &1& 0 \\ 0 & 0 &1 \end{array}\right],\label{B12}\\
&&B_{23}=B_{2}=\frac{\textrm{i}\omega}{\sqrt{3}}\left[\begin{matrix}
\omega & 0 & 0 & 0 & 0 & \omega & 0 & 1 & 0\\
0 & \omega & 0 & \omega & 0 & 0 & 0 & 0 & 1\\
0 & 0 & \omega & 0 & \omega & 0 & 1 & 0 & 0\\
0 & 1 & 0 & \omega & 0 & 0 & 0 & 0 & \omega\\ 
0& 0 & 1 & 0 & \omega & 0 & \omega & 0 & 0\\
1 & 0 & 0 & 0 & 0 & \omega & 0 & \omega & 0\\
0 & 0 & \omega & 0 & 1 & 0 & \omega & 0 & 0\\
\omega & 0 & 0 & 0 & 0 & 1 & 0 & \omega & 0\\
0 & \omega & 0 & 1& 0 & 0 & 0 & 0 & \omega
\end{matrix}\right],\label{B23}
\end{eqnarray}
obeying 
\begin{equation}\label{B123}
B_{12}B_{23}B_{12}=B_{23}B_{12}B_{23}.
\end{equation}
Here we emphasize that the braid relation in Eq.(\ref{B123}) and YBE is expressed in terms of three nearest parafermions, which only occupy 2-qutrit space $(\mathbb{C}_3)^{\otimes2}$ totally. Under the rational Yang-Batxerization of braid matrix, the unitary $\breve{R}$-matrix satisfying YBE shown in Eq. (\ref{YBEcondition}) can be obtained \cite{yu2016z3}
\begin{equation}\label{9solution1}
\breve{R}_{i}(\theta)=\frac{2}{\sqrt{3}}(\cos(\theta+\pi/6)+\sin\theta B_i).
\end{equation}
 which reduces braid matrix at $\theta=\pi/3$ and corresponds to the maximal von Neumann entropy of qutrit states we shall show below. 
Here we note that the YBE in Eq. (\ref{YBEcondition}) is independent of the concrete representation of $\breve{R}(\theta)$, but the relation of three parameters is related to the $\breve{R}$. Under the solution shown in Eq. (\ref{9solution1}), the constraint from YBE leads to the condition for three parameters \cite{yu2016z3}
\begin{equation}\label{YBEangularrelation}
\tan\theta_2=\frac{\tan\theta_1+\tan\theta_3}{1+\frac{1}{3}\tan\theta_1\tan\theta_3}.
\end{equation}
Different from the traditional parameter relation $u_2=u_1+u_3$(Galilean type additivity) in Yang-Baxter equation in the usual chain models \cite{faddeev1982integrable,kulish1981lecture,korepin1997quantum}, the new relation for three parameters in Eq. (\ref{YBEangularrelation}) obeys Lorentzian type additivity $u_2=\frac{u_1+u_3}{1+u_1u_3}$ for $u_1=\frac{\tan\theta_1}{\sqrt{3}}, u_2=\frac{\tan\theta_2}{\sqrt{3}}, u_3=\frac{\tan\theta_3}{\sqrt{3}}$.
\section{$\ell_1$-norm and von Neumann entropy in Yang-Baxter equation }\label{Sec3}
In this section, we shall show that the maximal 2-qutrit von Neumann entropy is determined by the braid matrix from parametrized $\breve{R}(\theta)$-matrix, i.e. $\theta=\pi/3$ for YBE. The braid matrix can also be obtained by extremization of $\breve{R}$-matrix.

We first obtain the reduced $3\times3$ $\breve{R}$-matrix as well as braid matrix from the original $9\times9$ ones. Under the 2-qutrit orthonormal basis $\{|1\rangle,|2\rangle,|3\rangle\}\otimes\{|1\rangle,|2\rangle,|3\rangle\}$, where $|1\rangle=(1,0,0)^T,|2\rangle=(0,1,0)^T,|3\rangle=(0,0,1)^T$, the $\breve{R}$-matrix can be expressed by the ket-bra representation, as
\begin{eqnarray}
&&\begin{split}
\breve{R}_{12}=&e^{-i\theta}|11\rangle\langle11|+e^{i\theta}|23\rangle\langle23|+e^{-i\theta}|32\rangle\langle32|\\
                         &+e^{-i\theta}|12\rangle\langle12|+e^{i\theta}|21\rangle\langle21|+e^{-i\theta}|33\rangle\langle33|\\
                         &+e^{-i\theta}|13\rangle\langle13|+e^{i\theta}|22\rangle\langle22|+e^{-i\theta}|31\rangle\langle31|.
\end{split}\\
&&\begin{split}
\breve{R}_{23}=&(\cos\theta-\frac{i}{3}\sin\theta)\,\left(|11\rangle\langle11|+|23\rangle\langle23|+|32\rangle\langle32|\right)\\
                         &+\frac{2i}{3}\omega^2\sin\theta \,(|11\rangle\langle23|+|23\rangle\langle32|+|32\rangle\langle11|)\\
                         &+\frac{2i}{3}\omega\sin\theta \,(|11\rangle\langle32|+|23\rangle\langle11|+|32\rangle\langle23|)\\
                         &+(\cos\theta-\frac{i}{3}\sin\theta)\,\left(|12\rangle\langle12|+|21\rangle\langle21|+|33\rangle\langle33|\right)\\
                         &+\frac{2i}{3}\omega^2\sin\theta \,(|12\rangle\langle21|+|21\rangle\langle33|+|33\rangle\langle12|)\\
                         &+\frac{2i}{3}\omega\sin\theta \,(|12\rangle\langle33|+|21\rangle\langle12|+|33\rangle\langle21|)\\
                         &+(\cos\theta-\frac{i}{3}\sin\theta)\,\left(|13\rangle\langle13|+|22\rangle\langle22|+|31\rangle\langle31|\right)\\
                         &+\frac{2i}{3}\omega^2\sin\theta \,(|13\rangle\langle22|+|22\rangle\langle31|+|31\rangle\langle13|)\\
                         &+\frac{2i}{3}\omega\sin\theta \,(|13\rangle\langle31|+|22\rangle\langle13|+|31\rangle\langle22|).                  
\end{split}
\end{eqnarray}
Under the unitary transformation $U$,
\begin{equation}
U=\left[\begin{matrix}
1 & 0 & 0 & 0 & 0 & 0 & 0 & 0 & 0\\
0 & 0 & 0 & 1 & 0 & 0 & 0 & 0 & 0\\
0 & 0 & 0 & 0 & 0 & 0 & 1 & 0 & 0\\
0 & 0 & 0 & 0 & 1 & 0 & 0 & 0 & 0\\ 
0 & 0 & 0 & 0 & 0 & 0 & 0 & 1 & 0\\
0 & 1 & 0 & 0 & 0 & 0 & 0 & 0 & 0\\
0 & 0 & 0 & 0 & 0 & 0 & 0 & 0 & 1\\
0 & 0 & 1 & 0 & 0 & 0 & 0 & 0 & 0\\
0 & 0 & 0 & 0 & 0 & 1 & 0 & 0 & 0
\end{matrix}\right],
\end{equation} 
the $9\times9$ basis $\{|11\rangle, |12\rangle, |13\rangle, |21\rangle, |22\rangle, |23\rangle, |31\rangle, |32\rangle, |33\rangle\}$ is reordered to be $\{|11\rangle, |23\rangle, |32\rangle, |12\rangle, |21\rangle, |33\rangle, |13\rangle, |22\rangle, |31\rangle\}$. Then it is not difficult to find that the two $\breve{R}$'s are reducible and can be expressed as the direct sum of three $3\times3$ matrices in three 3-D subspaces, $\{|11\rangle, |23\rangle, |32\rangle\}$, $\{|12\rangle, |21\rangle, |33\rangle\}$, $\{|13\rangle, |22\rangle, |31\rangle\}$, respectively. Indeed, the three subspaces can be categorized by the 2-qutrit ``parity operator'' defined as follows. Introducing the 2-qutrit ``parity operator'' $P$,
\begin{equation}\label{parity}
P^3=1,\, P|ij\rangle=\omega^{(i+j)}|ij\rangle,\quad \omega=\exp(i2\pi/3).
\end{equation}
It is easy to check that $P=\omega^2$ in the subspace $\{|11\rangle, |23\rangle, |32\rangle\}$, $P=1$ in the subspace $\{|12\rangle, |21\rangle, |33\rangle\}$, $P=\omega$ in the subspace $\{|13\rangle, |22\rangle, |31\rangle\}$, and $\breve{R}$-matrix commutes with the parity $P$,
\begin{equation}
[\breve{R}_{12},P]=[\breve{R}_{23},P]=0.
\end{equation} 
That is to say, the $\breve{R}$ operation(including braid operation) preserves the 2-qutrit parity $P$ of the system. Taking one subspace $\{|11\rangle, |23\rangle, |32\rangle\}$ as example, the reduced $3\times3$ $\breve{R}$-matrix reads
\begin{eqnarray}
&&\mathcal{A}_{12}(\theta)=\left[\begin{array}{ccc} e^{-i\theta} & 0 & 0 \\ 0 & e^{i\theta} & 0 \\ 0 & 0 & e^{-i\theta} \end{array}\right],  \\
&&\mathcal{A}_{23}(\theta)=\left[\begin{smallmatrix}\cos\theta-\frac{i}{3}\sin\theta &\frac{2i}{3}\omega^2\sin\theta& \frac{2i}{3}\omega\sin\theta \\ \frac{2i}{3}\omega\sin\theta & \cos\theta-\frac{i}{3}\sin\theta & \frac{2i}{3}\omega^2\sin\theta  \\ \frac{2i}{3}\omega^2\sin\theta & \frac{2i}{3}\omega\sin\theta & \cos\theta-\frac{i}{3}\sin\theta \end{smallmatrix}\right].             
\end{eqnarray}
They satisfy YBE 
\begin{equation}
\mathcal{A}_{12}(\theta_1)\mathcal{A}_{23}(\theta_2)\mathcal{A}_{12}(\theta_3)=\mathcal{A}_{23}(\theta_3)\mathcal{A}_{12}(\theta_2)\mathcal{A}_{23}(\theta_1),
\end{equation}
with the same constraint as pointed in Eq.(\ref{YBEangularrelation})
\begin{equation}
\tan\theta_2=\frac{\tan\theta_1+\tan\theta_3}{1+\frac{1}{3}\tan\theta_1\tan\theta_3}.
\end{equation}

Now we introduce how to screen out the braid relation from YBE by extremizing $\ell_1$-norm and von Neumann entropy of the state generated by $\breve{R}$-matrix . Acting $\mathcal{A}_{23}(\theta)$ on basis $|11\rangle$, one obtains the state in the subspace $\{|11\rangle, |23\rangle, |32\rangle\}$
\begin{equation}
|\Psi(\theta)\rangle=(\cos\theta-\frac{i}{3}\sin\theta)|11\rangle+\frac{2i}{3}\omega^2\sin\theta|32\rangle+\frac{2i}{3}\omega\sin\theta|23\rangle.
\end{equation}
Following the definition in Eq. (\ref{l1norm}), the $\ell_1$-norm of the state $|\Psi(\theta)\rangle$ is expressed as
\begin{equation}
||\Psi||_{\ell_1}=|\cos\theta-\frac{i}{3}\sin\theta|+|\frac{2i}{3}\omega^2\sin\theta|+|\frac{2i}{3}\omega\sin\theta|.
\end{equation}

\begin{figure}[!htb]
\includegraphics[width=0.8\columnwidth,clip]{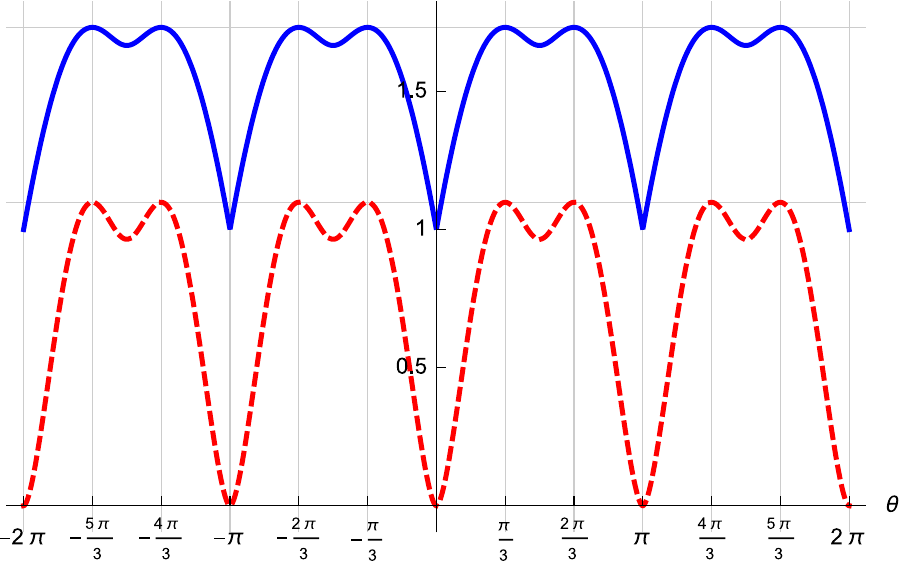}
\caption{Von Neumann entropy and $\ell_1$-norm of $|\Psi(\theta)\rangle$ as a function of $\theta$ that describes the qutrit entanglement. The von Neumann entropy is labeled by the red dashed line, and the $\ell_1$-norm is labeled by blue solid line. Both of them achieve the maximal extremum at $\theta=\frac{\pi}{3}$, at which value the $\breve{R}(\theta)$ reduces braid matrix.}\label{L1vN}
\end{figure}

On the other hand, the von Neumann entropy \cite{nielson2010quantum}, which reflects the quantum entangled degree of the two subsystems, reads
\begin{equation}
S(\rho)=-\tr[\rho \ln\rho],
\end{equation}
where $\rho$ represents the partial trace of the  density matrix of the system. In our case with the quantum state $|\Psi(\theta)\rangle$, the von Neumann entropy is expressed by $\theta$, as
\begin{equation}
\begin{split}
S=&-|\cos\theta-\frac{i}{3}\sin\theta|^2\ln|\cos\theta-\frac{i}{3}\sin\theta|^2\\
&-|\frac{2i}{3}\sin\theta|^2\ln|\frac{2i}{3}\sin\theta|^2-|\frac{2i}{3}\sin\theta|^2\ln|\frac{2i}{3}\sin\theta|^2.
\end{split}
\end{equation}
The values of $\ell_1$-norm and von Neumann entropy $S$, as functions of parameter $\theta$ in $\breve{R}$-matrix,  are shown in Fig. \ref{L1vN}. From Fig. \ref{L1vN} we conclude that the location of the extremum of $\ell_1$-norm and von Neumann entropy of $|\Psi(\theta)\rangle$ coincide each other exactly. Especially, the two quantities both achieve the maximal values  at $\theta=\frac{\pi}{3}$, meanwhile the $\breve{R}(\theta)$ turns to be braid matrix. The result is also applied to the qubit cases, which has been discussed in Ref.  \cite{niu2011role}.  Hence we conclude that the extreme values of the two quantities $||\Psi||_{\ell_1}$ and $S$ endow the real physical meaning for $\breve{R}$-matrix.
\section{qutrit realization of 4-anyon topological basis and  reduced $2\times2$ $\breve{R}$-matrix }\label{Sec4}
In this section, we give an entangled 4-qutrit representation of 4-anyon topological basis. By acting the localized $9\times9$ unitary representation of T-L elements on the topological basis, we obtain the $2\times2$ representation of T-L algebra, which exactly corresponds to the the Jones representation of the braid groups obtained from diagrammatic topological fusion basis.

To start with, let us briefly review some basic knowledge about the graphic representation of  braiding, Temperley-Lieb algebra and topological fusion basis\cite{kauffman2001knots,wang2010topological}.

Graphically, the T-L algebra reads
\begin{equation*}
\ununun=\unL, \quad\ \ \UNUN=\UON=\smyloop \ \Un,
\end{equation*}
\begin{equation}
  T_i = \Unij ,\quad d=\myloop\ (\text{value of loop}).
\end{equation}
The braid operation $\cross$ in Eq. (\ref{BTL}) can be decomposed into the combination of identity operator and  T-L element  under the Skein relation, graphically, as
\begin{equation}
 \cross=-e^{\frac{i\pi}{12}}\left(\alpha\llll+\alpha^{-1}\Un\right), \, d=\myloop,
\end{equation}
where $\alpha=ie^{\frac{i\pi}{12}}$, and quantum dimension $d=-\alpha^2-\alpha^{-2}=\sqrt{3}$. In $SO(3)_2=SU(2)_4$ Chern-Simons(CS) theory, the dimension of Hilbert space for 4-anyon fusion states is two \cite{levaillant2016topological,hikami2008skein}, with each anyon has topological charge 1.  Under the Jones-Wenzl idempotent \cite{jones1983index,wenzl1987on}, the two orthonormal fusion basis can be expressed as cups
\begin{equation}\label{topological basis}
\begin{split}
   &|e_1\rangle =  \frac{1}{d}\, \Uu,\\
   &|e_2\rangle = \frac{1}{\sqrt{d^2-1}}\left( \twojoin -\frac{1}{d} \Uu \right).
\end{split}
\end{equation}
Acting the T-L elements on the above two basis $\{|e_1\rangle, |e_2\rangle\}$, one obtains 
\begin{equation}\label{GTL2}
\begin{split}
&T_{12}=T_1=\left[\begin{array}{cc} \sqrt{3} & 0  \\ 0 & 0  \end{array}\right],  \\
&T_{23}=T_2=\frac{1}{\sqrt{3}}\left[\begin{array}{cc} 1 & \sqrt{2}  \\  \sqrt{2} & 2 \end{array}\right], \\
&T_{34}=T_3=\left[\begin{array}{cc} \sqrt{3} & 0  \\ 0 & 0  \end{array}\right],
\end{split}
\end{equation}
which correspond to the Jones representation of the braid group \cite{wang2010topological}. 

In the previous paper \cite{niu2011role}, the $4\times4$ localized unitary representation of braid operator with quantum dimension $d=\sqrt{2}$ has been connected with the Jones representation of braid under the qubit realization of 4-anyon fusion basis for $SU(2)_2$ Chern-Simons theory. We now extend the result to $9\times9$ metaplectic anyon system with quantum dimension $d=\sqrt{3}$ associated with the $SO(3)_2=SU(2)_4$. 

Due to the simple relation between braid operators and T-L elements as shown in Eq. (\ref{BTL}), we consider the T-L elements directly. The $9\times 9$ localized unitary representation of T-L algebra in $(\mathbb{C}_3)^{\otimes N}$ reads ($I$ is $3\times3$ identity matrix) \cite{rowell2012localization, yu2016z3} 
\begin{equation}\label{LURT}
\begin{split}
&T'_i=I\otimes I\cdots\otimes \underset{i,i+1}{T'}\otimes\cdots I\otimes I,\\
&T'=\frac{1}{\sqrt{3}}\left[\begin{matrix}
1 & 0 & 0 & 0 & 0 & \omega & 0 & \omega^2 & 0\\
0 & 1 & 0 & 1 & 0 & 0 & 0 & 0 & 1\\
0 & 0 & 1 & 0 & \omega^2 & 0 & \omega & 0 & 0\\
0 & 1 & 0 & 1 & 0 & 0 & 0 & 0 & 1\\ 
0& 0 & \omega & 0 & 1 & 0 & \omega & 0 & 0\\
\omega^2 & 0 & 0 & 0 & 0 & 1 & 0 & \omega & 0\\
0 & 0 & \omega^2 & 0 & \omega & 0 & 1 & 0 & 0\\
\omega & 0 & 0 & 0 & 0 & \omega^2 & 0 & 1 & 0\\
0 & 1 & 0 & 1& 0 & 0 & 0 & 0 & 1
\end{matrix}\right].
\end{split}
\end{equation}
Under the representation of $\mathbb{Z}_3$ parafermions shown in Eq. (\ref{z3jw}), $T'_i$ can be reexpressed by Eq. (\ref{z3para}), as
\begin{equation}\label{TLE2}
T'_i\equiv\frac{1}{\sqrt{3}}\left(1+\omega^2C_{2i-1}^{\dagger}C_{2i+2}+\omega^2C_{2i-1}C_{2i+2}^{\dagger}\right).
\end{equation}
Since the braid matrix and T-L matrix are closely related to the 2-qutrit maximal entangled states defined by maximal von Neumann entropy, it is reasonable to express the T-L elements by 2-qutrit entangled states with maximal entangled states. Considering the 2-qutrit parity symmetry of the T-L matrix, we introduce three maximal  entangled states, ($i,j \in\{1,2,3,4\}$)
\begin{eqnarray}
&&|\alpha_{ij}\rangle=\frac{1}{\sqrt{3}}\left(|11\rangle+\omega^2|23\rangle+\omega|32\rangle\right)_{ij},\\
&&|\beta_{ij}\rangle=\frac{1}{\sqrt{3}}\left(|12\rangle+|21\rangle+|33\rangle\right)_{ij},\\
&&|\gamma_{ij}\rangle=\frac{1}{\sqrt{3}}\left(|13\rangle+\omega|22\rangle+\omega^2|31\rangle\right)_{ij},
\end{eqnarray}
where $i,j\,(i\neq j)$  represent 2 different qutrit sites. 

It is interesting that the T-L elements $T'_i$ in Eq. (\ref{LURT}) can be expressed by the three maximal entangled states, as (j=i+1)
\begin{equation}\label{TLp}
T'_i=\sqrt{3}\left(|\alpha_{ij}\rangle\langle\alpha_{ij}|+|\beta_{ij}\rangle\langle\beta_{ij}|+|\gamma_{ij}\rangle\langle\gamma_{ij}|\right),\,(\textrm{j=i+1}).
\end{equation}
Moreover, the 4-anyon topological basis in Eq.(\ref{topological basis}) for $d=\sqrt{3}$ can also be represented by 4-qutrit states,
\begin{equation}\label{TBqutrit}
\begin{split}
&|e_1\rangle=\frac{1}{\sqrt{3}}\left(|\alpha_{12}\rangle|\gamma_{34}\rangle+|\beta_{12}\rangle|\beta_{34}\rangle+|\gamma_{12}\rangle|\alpha_{34}\rangle\right),\\
&\begin{split}
|e_2\rangle=&\frac{i}{\sqrt{2}}(\omega|\alpha_{23}\rangle|\gamma_{41}\rangle+|\beta_{23}\rangle|\beta_{41}\rangle+\omega|\gamma_{23}\rangle|\alpha_{41}\rangle)\\
&-\frac{1}{\sqrt{2}}|e_1\rangle.
\end{split}
\end{split}
\end{equation}
In the 4-qutrit space $(\mathbb{C}_3)^{\otimes 4}$, there are totally three T-L elements. Acting $T'_i$ in Eq. (\ref{TLp}) on the two basis in Eq. (\ref{TBqutrit}), one obtains
\begin{equation}\label{QTL2}
\begin{split}
&T'_{12}=E'_1=\left[\begin{array}{cc} \sqrt{3} & 0  \\ 0 & 0  \end{array}\right],\\
&T'_{23}=E'_2=\frac{1}{\sqrt{3}}\left[\begin{array}{cc} 1 & \sqrt{2}  \\  \sqrt{2} & 2 \end{array}\right],\\
&T'_{34}=E'_3=\left[\begin{array}{cc} \sqrt{3} & 0  \\ 0 & 0  \end{array}\right].
\end{split}
\end{equation}
Substituting Eq.(\ref{QTL2}) into Eq.(\ref{BTL}), we obtain the $2\times2$ braid matrices
\begin{equation}\label{QB2}
\begin{split}
&B'_{12}=B'_1=\left[\begin{array}{cc} e^{i\frac{\pi}{3}} & 0  \\ 0 & e^{-i\frac{\pi}{3}}  \end{array}\right],\\
&B'_{23}=B'_2=\frac{1}{\sqrt{3}}\left[\begin{array}{cc} e^{-i\frac{\pi}{6}} & i\sqrt{2}  \\  i\sqrt{2} & e^{i\frac{\pi}{6}} \end{array}\right],\\
&B'_{34}=B'_3=\left[\begin{array}{cc} e^{i\frac{\pi}{3}} & 0  \\ 0 & e^{-i\frac{\pi}{3}}  \end{array}\right].
\end{split}
\end{equation}
In comparison with Eq. (\ref{GTL2}), we find that the obtained $2\times2$ $B'_i$ are exactly the Jones representation of braid matrix in 4-strand space. Thus by constructing the parity-preserved qutrit entangled states with maximal von Neumann entropy and expressing the T-L elements and topological basis by the entangled states, we obtain the relationship between localized unitary representation and Jones representation of braid for 4-strand system. 
\section{Conclusion and Discussion}
This paper mainly includes two results about $\mathbb{Z}_3$ parafermion representation of braiding associated with quantum information. All of the results are related to reducing higher dimensional representation of braid matrices to the lower ones. The first is reducing the $9\times9$ braid matrices in Eq. (\ref{B12}) and Eq. (\ref{B23}) to $3\times3$ by introducing parity $P$-preserved subspace, such as $\{|11\rangle, |23\rangle, |32\rangle\}$ (One can also choose other two subspaces, which do not change the results). In the subspace, the braid operators with real physical meaning can be screened out  from $\breve{R}$-matrix with the aid of 2-qutrit maximal entangled states, where both $\ell_1$-norm and von Neumann entropy achieve their maximal values. The second is reducing the localized $9\times9$ unitary representation of braiding relating to T-L elements in Eq. (\ref{LURT}) into $2\times2$ Jones representation in 4-anyon basis. Under the representation, the behavior of localized braid matrix coincides well with the Jones representation of braiding, which can be obtained by braiding the fusion basis diagrammatically.  In this sense, the qutrit representation does reflect the braiding properties. However, how to propose a well defined topological entanglement entropy for fusion basis, as proposed in Ref. \cite{hikami2008skein}, by means of qutrit language is still in challenge. Hence, the result can only be regarded as a qutrit simulation of diagrammatic braiding operation for 4-strand system, where each strand corresponds to a sector with topological charge 1.

Indeed, both of the above two results rely partly on the maximal entangled states(defined by von Neumann entropy). These are due to the parity preservation ( see Eq. (\ref{parity})) of the braiding operation on the natural basis $|ij\rangle$. Concretely, for 2-qutrit pure states, the $9\times9$ Hilbert space can be classified into three $3\times3$ subspaces by eigenvalues of parity $P$. Taking one subspace $\{|11\rangle, |23\rangle, |32\rangle\}$  as example, where  $P=\omega^2$, the superposed state $|\psi\rangle=a_1|11\rangle+a_2|23\rangle+a_3|32\rangle$ achieves its maximal von Neumann entropy and $\ell_1$-norm simultaneously at equal distribution $|a_1|=|a_2|=|a_3|=1/\sqrt{3}$. It  not only reflects the role of parity $P$ plays in generating entanglement in the 2-qutrit Hilbert space, but also guarantees the $\ell_1$-norm and von Neumann entropy achieves their maximal values simultaneously. Some of the results can also be extended to $\mathbb{Z}_{N>3}$ parafermion solution of YBE, but for $N>4$, the rational Yang-Baxterized $\breve{R}$-matrix is not unitary.  


In conclusion, we take $\mathbb{Z}_3$ parafermion model as examples to show the important roles of the parity, $\ell_1$-norm and von Neumann entropy for 2-qutrit in determining the braiding behavior of Yang-Baxter equation.  How to make connections between N-qudit  entanglement and N-strand localized unitary $D^2\times D^2$ representation of braid group is still an open problem. On the other hand, extending the 4-qutrit realization of 4-anyon topological basis to n-anyon basis is also worthy of doing. 
\section{Acknowledgements}
This work is in part supported by NSF of China(Grants No. 11475088).

\bibliographystyle{unsrt}

\end{document}